\documentstyle[12pt]{article}
\begin{document}

\begin{flushright}
RU--97--44\\
\today 
\end{flushright}

\begin{center}
\bigskip\bigskip
{\Large \bf Modeling The Glueball Spectrum By A Closed Bosonic Membrane }
\vspace{0.3in}      

{\bf Gregory  Gabadadze}
\vspace{0.2in}

{\baselineskip=14pt
Department of Physics and Astronomy, Rutgers University \\
Piscataway, New Jersey 08855, USA}\\
{\rm email: gabad@physics.rutgers.edu }

\vspace{0.2in}
\end{center}

\vspace{0.9cm}
\begin{center}
{\bf Abstract}
\end{center} 
\vspace{0.2in}
We use an  analogy between the Yang-Mills theory Hamiltonian
and the matrix model description of the  closed bosonic membrane
theory  to calculate the spectrum of glueballs in the large $N_c$ limit. 
Some features of the Yang-Mills theory vacuum,  
such as  the screening of the 
topological charge and vacuum topological susceptibility
are discussed. We show that the topological susceptibility 
has different properties  depending on whether it is calculated in the
weak coupling or strong coupling regimes of the theory. 
A  mechanism of
the formation of the pseudoscalar glueball state  within 
pure Yang-Mills theory is proposed  and studied. 

\vspace{2cm}



bosonic membrane; pseudoscalar glueball.
\newpage
{\bf Introduction}
\vspace{0.1in} 

Gluodynamics \cite {GellMann}, being the asymptotically free 
theory \cite {AF}
of colored \cite {Color} massless vector particles, is believed 
to underline  the dynamics  of  strong interactions. 
Because of  asymptotic freedom  the theory 
is well studied  at short distances, however,  
long distance phenomena deserve to be understood much better. 

The theory predicts 
glueballs \cite {Glueballs}, the nonperturbative 
bound states composed of pure glue  
\cite {West}. That    prediction was   
confirmed some time ago  by ``observing'' 
glueballs in lattice QCD simulations  \cite {UKQCD}, \cite {Weingarten}.
In addition to that, there are  
experimental signatures  of resonances 
which strongly resemble properties of glueballs
(for a recent analysis of these issues see ref. \cite
{CFL}). 

Studying glueballs one might hope to learn more about the 
complicated ground state structure of non-Abelian 
Yang-Mills (YM) theory.
The main question one might wonder about is  the mechanism of
the formation of glueball  states in YM theory.
Those  states appear  to be heavy in comparison with the lightest  
hadrons and range, depending on the spin-parity structure,
within the mass interval $1.5-2.3~{\rm GeV}$ 
\cite {UKQCD}, \cite {Weingarten}. 
Thus, the naive picture of  the  glueball as a 
system of two massless gluons which interchange  
virtual perturbative gluons does not seem to be appropriate. 

In this work  we are looking for qualitative features responsible 
for the process of formation of pseudoscalar glueballs. 
A possible  mechanism will be proposed. 
As an outcome  we calculate  the spectrum of 
lightest pseudoscalar glueballs. The results  
are in agreement with predictions of lattice calculations \cite {UKQCD}.
The paper deals with pure YM theory, no light  fermions are included. 
A  brief discussion of full QCD is given at the end of the work.

Our study  
relies on  the existence of the $\theta$ term  in pure YM theory.   
We define the topological charge density operator as
$Q\equiv {1\over 32 \pi^2}
G^a_{\mu\nu}{\tilde G}^a_{\mu\nu}$, with $G^a_{\mu\nu}$   being the 
nonabelian gauge field  strength tensor and the dual tensor 
is normalized  as ${\tilde G}^a_{\mu\nu}=
{1\over 2}\varepsilon_{\mu\nu\alpha\beta}
G^a_{\alpha\beta}$. Because of instantons \cite {BPST},
the non-Abelian gauge theory possesses a  complicated vacuum structure
\cite {Vacuum}.
That is,  there is an infinite number of  degenerate vacuum states 
labeled by some topological  invariant, the winding number or
topological charge.
Instantons, being defined in Euclidean space,  provide that 
quasi-classical tunneling processes happen  between the different
vacua. Thus, the true ground state of the theory 
is  a  superposition
of the vacua with different topological charges. The   superposition
can be provided  in the path integral formulation by adding  
to the action  the $\theta$ term $\Delta 
S_{\theta}\equiv \theta \int d^4x Q(x)$ \cite {Vacuum}.
However, such  a modification is not
a harmless procedure. The $\theta$ term in QCD leads  to an  
induced neutron electric dipole moment.  
Experimental bounds on that  quantity  
restrict the value of the $\theta$ parameter
to be unnaturally small, less than the  billion'th part of the unity, 
and give rise the famous strong CP problem \cite {Peccei}. 

The picture outlined   above implies  
that the integral of the topological charge density $\int d^4x Q(x)$,
being the topological charge in Euclidean space,  is quantized
if the instanton boundary conditions are imposed on gauge fields
\footnote{Under those  boundary condition one means that  
the vector potential $A_\mu$ tends to a pure gauge configuration 
at spatial infinity $ A_\mu \rightarrow U^{-1}(x)\partial_\mu U(x)$,
with $U(x)$ being an  element of the $SU(N_c)$ gauge group.}. 
Thus, the  whole scenario of the superposition of the different
vacua by means of the $\theta$ term  relies 
on  quasiclassical arguments. In general, one expects that 
the quasiclassical approximation is  justified
in a  weak coupling regime only \cite {Coleman}.
What the $\theta$ term leads to in a  strong coupling
approximation where quasiclassical arguments cease to be valid, 
is not clear $a~priori$. 

It  was argued by Witten \cite{Witten1} that 
in the confining phase of the theory
noninteracting instanton boundary conditions should not be  
relevant. The key
observation was  that gauge fields with  instanton 
boundary conditions 
do not yield the area law for the Wilson loop. Thus,
configurations with those boundary conditions, as any
configurations tending to a pure gauge at infinity,  
fail to satisfy the confinement criterion  \cite {Wilson}. 
As a result, in
the strong coupling approximation of the  confining theory 
one should rather   encounter some 
smeared distributions  of interacting  topological  charges as  opposed 
to the noninteracting instanton  system  with  quantized topological
charge \cite {Witten1}. 
This  statement finds  support in recent models of the YM vacuum.
Properties of hadrons are correctly described by 
the model in which  instantons and antiinstantons 
are coupled  in moleculelike (or even more complicated) entities
\cite {Liquid}.
This kind of instanton clustering, indicating on strong correlations   
between them, was also observed  in some recent 
lattice QCD studies \cite {KovacsDeGrand}. 

Interactions between instantons, if sufficiently strong, 
lead to the screening of the topological charge 
at finite distances \cite {Samuel}, \cite {Screening}, 
in analogy with the   well-know  phenomenon in plasma physics.

A quantitative study  of 
the screening  phenomenon from  fundamental principles
is a cumbersome task. However, as we mentioned already,
both  hadron phenomenology and   lattice simulations 
seem to support that picture. 
Below  in Section 1 we  present some  
arguments (other than the ones mentioned above) 
indicating that the screening  of the topological
charge should really exist. We show (Section 2)
that the three-form composite field, which is Hodge dual to
the Chern-Simons current and is known to propagate
the Coulomb type interaction, should be relevant 
for the description of the screening phenomenon. 

In Section 2 we present a possible mechanism  of the   formation of 
a pseudoscalar glueball. The main ingredients needed 
for that mechanism to be realized in a confining theory are 
the screening of the topological
susceptibility and the presence  of the $\theta$ angle in the theory. 

Having this part set, 
we discuss in Section 3  an  analogy  between the spectrum of the YM
Hamiltonian and that of a closed bosonic membrane with the topology
of a sphere. We use that analogy to calculate the spectrum of 
glueballs. In fact, we derive a  matching condition 
between the spectrum  of YM theory in the large $N_c$ limit 
and the matrix quantum mechanics
formulation  of the closed bosonic membrane theory. 
Then, studying the spectrum of a spherical closed membrane
and using the matching condition we calculate the spectrum of the 
YM Hamiltonian. That  gives the prediction for masses
of lightest glueballs. We also show  that the scenario discussed
in this work is realized only when the $\theta$ parameter  
is a macroscopic number, 
i.e. a  number of order of the unity or so. 
More accurate estimates are given in
Subsection 3.3. We briefly discuss how the strong CP violation, being 
present in pure YM theory, might still not be observable in 
full QCD with light quarks. 

Discussions in  the present paper are based  on 
the results obtained by M. L\"uscher \cite {LuscherSus}, 
\cite {Luscher}, \cite {LuscherNP},  by J. Goldstone and J. Hoppe 
(see ref. \cite {Hoppe1}), 
and by B. de Wit, J. Hoppe and H. Nicolai \cite {Hoppe2}. Where
it is possible we present  below brief summaries  of those results. 
\vspace{0.2in} \\
{\bf 1. Topological Susceptibility}
\vspace{0.1in} 

In this section we study  properties of the 
correlator of the vacuum topological susceptibility. 
We work in Euclidean space-time
assuming  that the theory is defined in a  compact Euclidean 
four-volume $V\equiv l^3\times \tau$, with  $l$ being  the linear  size
of the volume and $\tau$ stands  for Euclidean time. 
The correlator of the vacuum topological
susceptibility $\chi(V)$ can be written  as follows: 
\begin{eqnarray}
\chi(V)\equiv \int_V \langle 0|T Q(x) Q(0)|0\rangle d^4x,
\end{eqnarray}
where $Q$ is the topological charge density operator defined   in the 
previous section\footnote[5]{ 
The correlator (1) is in general a divergent
Green's function. These divergences can be removed by means 
of standard procedures which are discussed in the Appendix.}. 

The function
$\chi(V)$ is determined  by the nonperturbative sector  
of Yang-Mills theory.  Calculation of $\chi(V)$ in that respect 
is a matter of  modeling  of  the  vacuum structure 
of  Yang-Mills theory. This in its turn is 
a complicated task. Below  we show  however,  that 
one can still study qualitative features of the 
volume dependence of the correlator of the 
vacuum topological susceptibility for  small and 
for large values of the volume\footnote[1]{We vary $l$ keeping $\tau$
fixed. Thus, we actually study the three-volume dependence
of $\chi(V)$.}. 
Let us first define, following \cite {LuscherSus},  
what one could  call a small volume or large volume limit.

There is a dynamically generated mass scale in YM theory, 
$\Lambda_{\rm YM}$.
The reciprocal  quantity of $\Lambda_{\rm YM}$ sets the characteristic
correlation length for the model. Let us denote that length 
by $\zeta\equiv \Lambda_{\rm YM}^{-1}$. 
Restricting for simplicity the YM beta function 
to  the next-to-leading order
approximation, 
the expression for $ \zeta $ can be written as follows:\footnote[2]{Since we
are interested in the magnitude of this quantity the 
explicit $ \theta $ dependence in this expression is dropped.} 
$$
\zeta=\mu^{-1} (\alpha_s)^{{\beta_1\over 2 \beta_0^2}}
~\exp \left ({2 \pi \over \beta_0 \alpha_s} \right ),
$$
where $\mu$ is the renormalization scale, 
$\alpha_s\equiv \alpha_s(\mu^2/\Lambda_{\rm YM}^2)$ is the 
scale-dependent strong
coupling constant, and $\beta_0$ and $\beta_1$
are the first two scheme independent  coefficients of the beta
function, $\beta_0=11 N_c/3$, $\beta_1=34 N_c^2/3$. The expression for
$ \zeta $ is renormalization group invariant in the corresponding order. 

Let us now introduce the following two limits. One 
can define the value of the volume element 
$V$ to be small if the correlation length $\zeta$ is much larger than  
$V^{1/4}$, i.e.,  $\zeta >>V^{1/4}$ \cite {LuscherSus}. 
The large volume limit in that case would 
refer to a volume element satisfying  the condition  $\zeta << V^{1/4}$.

One can show 
that the two limits defined above correspond respectively to
the weak coupling and strong  coupling regimes of the theory. 
In order to see this let us keep the product of 
the renormalization scale $\mu$ and the value of $V^{1/4}$ fixed, say,
$V\mu^4=1.$  This condition sets the scale $\mu$ as an  infrared cutoff.
Then, the expression for the correlation length
takes  the form
$$
{\zeta \over  V^{1/4}}= (\alpha_s)^{\beta_1\over 2 \beta_0^2}
\exp \left ({2 \pi \over \beta_0 \alpha_s} \right).
$$
Thus,  the small volume approximation given by the condition  
$\zeta >> V^{1/4}$  corresponds
to the weak coupling regime, i.e.,  $\alpha_s << 1$. For instance,
if  one sets ${\zeta \over  V^{1/4}}\simeq 10>>1$, the corresponding
value of the coupling constant is  $\alpha_s\simeq 0.2$\footnote[4]
{If one approximates $\Lambda_{\rm YM}\simeq (100-200)~{\rm MeV}$,
then the small volume limit refers to $V^{1/4}<<(1-2)~{\rm fm}$.}. 

Let us now  turn to the discussion of the large volume limit defined
by the condition 
$\zeta << V^{1/4}$. This  limit is equivalent to
the  small $\mu$ approximation. However, for small values of $\mu$
the running coupling constant $\alpha_s$ is  a big number. Hence, 
the large volume limit corresponds to the strong coupling regime of
the theory.  Regretfully,  we can not estimate 
(as  we did in the case of the 
small coupling constant) how large the coupling constant should
actually be. 
The approximation which is used 
defining   $\zeta$
breaks down for large values  of the coupling. 
Though one could  present  the definition of $\zeta$ for  any orders
of perturbation theory (see, for example, \cite {Collins}), 
that   definition would contain  the exact form of the beta 
function $\beta(\alpha_s)$
which is known only perturbatively. 
So, the all order formula  also 
becomes inappropriate for practical calculations
in the strong coupling approximation. 

Below we show, at least qualitatively,  that 
the topological susceptibility as a function of $V$  
has different behavior depending
whether it is calculated in the weak or strong 
coupling regimes.  In the weak coupling phase 
it is an  increasing  function of the argument, and on the  contrary,  in the
strong coupling regime the function  decreases with the argument
monotonically. 
\vspace{0.2in} \\
{\it 1.1. The Weak Coupling Approximation}
\vspace{0.1in}

Let us start with the small volume  or weak coupling approximation.
The non-Abelian
gauge theory provides  a good description of physics in that domain. 
Excitations with a zero topological charge do 
not contribute to the 
value of $\chi (V)$ defined in eq. (1). Only nontrivial topological
configurations of gauge fields are to be taken into account.
In the weak coupling regime 
the YM vacuum can be 
approximated by noninteracting, well-separated  
instantons \cite {CDG77}.
In that  approximation  instantons can be treated 
as point-like objects.
The expression  for the topological charge density can be
written as follows: 
$$
Q(x)=\sum_i q_i \delta^{(4)}(x-x_i),
$$ 
where $q_i$ denotes the topological charge for a configuration
localized at the point $x_i$. Assuming that instantons do not 
interact with one another we derive
\begin{eqnarray}
\langle 0|TQ(x) Q(0)|0\rangle ={1\over V}
\sum_{q=-\infty}^{+\infty} q^2 P_q(V)~\delta^{(4)}(x), 
\end{eqnarray}
where the index $i$ in the definition of the topological charge was omitted.
The quantity $P_q$ denotes the probability 
for a nonabelian gauge field configuration to have 
a topological charge equal to $q$. 
These probabilities   are exponentially
suppressed for nonzero $q$ and one expects that the infinite series
in eq. (2) converges\footnote[8]{
The partition function $Z(\theta)$ in that case can be approximated as 
$Z(\theta)=\sum_q P_q \exp (i\theta q)$.}. 

Substituting  eq. (2) into eq. (1) one derives
\begin{eqnarray}
\chi(V)={1\over V}
 \sum_{q=-\infty}^{+\infty} q^2 P_q(V).
\end{eqnarray}
In the approximation we set above  the following relation is 
valid: 
\begin{eqnarray}
P_q(V)=(P_1(V))^{|q|},
\nonumber
\end{eqnarray}
where $P_1(V)$ denotes the one instanton contribution.
Substituting this  relation   into
eq. (3) and performing the summation of the infinite series  
one gets
the following expression for the topological susceptibility:
\begin{eqnarray}
\chi(V)={2 P_1(V)\over V}{1+P_1(V)\over (1-P_1(V))^3}.
\end{eqnarray}
Thus, the small volume behavior of the function $\chi(V)$ 
is approximately defined by eq. (4).
The expression for $P_1(V)$ can be calculated in the one loop
approximation using the 
well known results of ref. \cite {tHooft}
$$
P_1(V)={\rm const}\times 
\exp \left ( -{2\pi\over \alpha_s
(V^{1/2} \Lambda_{\rm YM}^2) } \right ), 
$$
where the following expression for the strong coupling
constant is supposed to be used 
$$
\alpha_s(V^{1/2} \Lambda_{\rm YM}^2)=-{4 \pi \over
\beta_0 {\rm ln}(V^{1/2} \Lambda_{\rm YM}^2)}+\dots
$$
The result for $P_1(V)$ is 
$$
P_1(V)={\rm const}\times  \left ({V\over \zeta^4} \right )^{\beta_0\over 4}
\times {\rm logarithms},
$$
where the logarithmic corrections  appear in the 
next-to-leading approximation. 
Hence, as $V\rightarrow 0$ the ratio $P_1(V)\over V$ also tends to
zero. As a consequence, in the small volume limit 
$\lim_{V\rightarrow 0}\chi (V)\rightarrow 0$.
Moreover, based on the relations given above 
one concludes  that  for small volume elements the quantity  
$\chi(V)$ is a  monotonically increasing function
of the argument  $V$.  This  property 
should hold as the condition $\zeta >> V^{1/4}$ is satisfied.

Suppose now that the quantity $V^{1/4}$ becomes comparable in
magnitude   
with  $\zeta$ so that  the weak coupling
approximation breaks down. As a result,  the 
pointlike noninteracting instanton approximation ceases  to be  valid.
Interactions  between instantons start to play a crucial role 
providing the screening of the topological charge \cite{Screening}.

Let us assume  for a moment  that one neglects  instanton 
interactions even for 
large values of the volume 
and let us study what happens in this unrealistic case keeping in
mind that the interaction effects are going to be included 
later.
Doing so one is  dealing with an ideal gas of instantons 
placed in a large volume. 
The approximate calculation of the partition function with
noninteracting instantons in the thermodynamic  limit yields
the following Gaussian  distribution 
function for $P_q(V)$ \cite{DMbig} 
\begin{eqnarray}
P_q(V >> \zeta^4) \approx  {1\over \sqrt {2\pi V d}} \exp 
\left ( -{q^2\over 2  V d} \right ), \nonumber
\end{eqnarray}
where $d$ is a not yet defined  positive constant. 
We substitute  this expression into eq. (3) and perform
the summation of the infinite series in the large volume limit. 
The final expression can be found 
using  the following relation
$$\sum_{q=-\infty}^{+\infty}e^{-\pi b q^2}= {1\over \sqrt b} 
\sum_{q=-\infty}^{+\infty}e^{-\pi q^2/b},$$  where $b$ is  an arbitrary 
positive number \cite {Macrobert}.
As a result one gets  
\begin{eqnarray}
\lim_{V >> \zeta^4 }~\chi (V)=d. 
\end{eqnarray}

Let us summarize briefly what we learned about the volume dependence 
of the topological susceptibility $\chi(V)$. 
In the zero  volume approximation the topological susceptibility
was zero. Increasing the value of $V$, so that the weak coupling
approximation still holds,   the function $\chi(V)$
increases monotonically. If one goes further and neglects 
the interaction between instantons even in the large volume 
(strong coupling)
approximation, one finds that the function $\chi(V)$ reaches its 
asymptotic value\footnote{These   properties were originally
studied  in ref. \cite {LuscherSus} considering
YM theory on a four-sphere $S^4$.}
denoted above by $\chi (V >> \zeta^4)=d$.
However, as we stressed earlier, interactions  between 
instantons play a crucial role
in the strong coupling
approximation. 
In the next subsection we show that
the topological susceptibility becomes 
a decreasing function of the argument for large values of $V$
when the effects of finite distance correlations between
topological charges are taken into account. 
\vspace{0.2in} \\
{\it 1.2. The Strong Coupling Approximation}
\vspace{0.1in}

Let us consider  the large volume  
or strong coupling limit.
In that limit the theory is in a confining phase.  
Instantons are interacting strongly\footnote{It is not even clear
whether it makes sense to talk about a configuration with a definite
topological charge in this case \cite {Witten1}.}. 
Those  interactions  become
responsible for formation of spin zero glueball states 
\cite {ShuryakGlueballs}.
A description in terms of colored variables is not a good approximation
anymore.
The theory, however, can be defined  by means of  
the low-energy effective
action containing colorless degrees of freedom.
The explicit form of that effective action for pure YM  theory is not known. 
In general, the  action
can be written as 
$$
S_{eff}=\int d^4x {\cal L}(G_n, \nabla G_n, \nabla ^2 G_n,...),
$$
where $G_n$'s  stand for  glueball fields.

Below we  study 
properties of the correlator of the vacuum topological susceptibility
in the  effective theory. We denote this quantity by $\chi_{\rm eff}(V)$.
The correlator in eq. (1)
is  saturated by the set of intermediate  
glueball  states 
\begin{eqnarray} 
\langle 0|T Q(x) Q(0)|0\rangle=d~ \delta^{(4)}(x)+\sum_n
\langle 0|Q|n\rangle  \langle n|Q|0\rangle    D_F(m_n |x|),
\end{eqnarray}
where $m_n$ is the mass of the $n$ th intermediate physical  state
and $D_F(m_n |x|)$ stands for the Euclidean $x$-space Feynman 
propagator of a  scalar massive particle
\begin{eqnarray}  
D_F(m_n |x|)={m_n\over 4\pi^2 |x|} K_1(m_n |x|),
\nonumber
\end{eqnarray}
with $K_1(m_n |x|)$ being the Bessel function of an imaginary argument. 

The parameter $d$ given in eq. (6) is a positive number.
It was introduced in the preceding section and 
in the simplest  case of a  dilute instanton gas approximation 
corresponds to the value of the topological 
susceptibility  in the large  volume limit. 
From the point of view of the effective theory we deal with, 
the parameter $d$ is a  momentum independent subtraction
coefficient in the dispersion relation for $\chi_{\rm eff}$ 
written in  momentum
space\footnote[2]{There is  another subtraction term in eq. (6). 
It is   proportional to the second derivative
of the Dirac delta function. 
This term,  being integrated in eq. (1) 
gives a  vanishing contribution  and 
does  not appear in the definition of $\chi(V)$. 
A  detailed discussion is given 
in the  Appendix.}. In eq. (6) we implicitly assumed that 
the volume element is sufficiently large so that the YM topological 
susceptibility occurring as the first term on the r.h.s. 
equals to its asymptotic value $d$. 

Strictly speaking, there are additional continuum contributions on the
r.h.s.  of eq. (6). They   account for possible many-particle 
intermediate states. 
Those contributions are studied 
in the Appendix.  We just mention here that the 
continuum  contributions do not affect the physical picture 
we are going to discuss in this subsection. 

One notices that  eq. (2), which includes only noninteracting instanton
effects,  reflects the lack  of 
finite distance correlations between topological charge densities,
i.e. the r.h.s. of eq. (2) is zero for any nonzero value of $x$. 
This would not be  the case  if  instanton interactions
were taken into account.
We also saw that the  insertion  of the intermediate glueball  states into  
eq. (2)  yields the expression (6) 
with finite distance correlations occurring on its r.h.s. 
Thus, one argues that the strong correlations between instantons, 
which are responsible for finite distance effects, 
are   phenomenologically included in eq. (6) as  
the  intermediate glueball states 
are taken into account.
The  argument above becomes more sensible if one recalls
that interactions  between instantons are responsible for the 
formation of those intermediate glueballs \cite {ShuryakGlueballs}. 

Let us now define the matrix elements occurring in eq. (6).
The operator of the topological charge density $Q$ is
an antihermitian operator in Euclidean space. 
Taking this into account one  introduces
the following parametrization for the matrix elements
\begin{eqnarray}
\langle 0|Q|n\rangle = -i f_n m_n^2,~~~~~~~~
\langle 0|Q|n\rangle  \langle n|Q|0\rangle  =
-f_n^2 m_n^4,
\nonumber
\end{eqnarray}
where $f_n$ can be thought of  as a decay constant of the corresponding
$n$'th glueball state (in analogy with  the pion decay constant $f_\pi$).
If one  substitutes  these definitions back into  eq. (6) the
following expression emerges 
\begin{eqnarray} 
\langle 0|T Q(x) Q(0)|0\rangle=d~
 \delta^{(4)}(x)-\sum_n
f_n^2 m_n^4 {m_n\over 4\pi^2 |x|}K_1(m_n |x|). 
\end{eqnarray}
Having this  relation  established let us study what happens with the
correlator of the topological susceptibility (eq. (1)). 
Substituting eq. (7)  into  eq. (1) we find 
\begin{eqnarray}
\chi_{\rm eff}(V)=d -\sum_{n}f_n^2m_n^2 ~ {\cal G}_n(V),  
\end{eqnarray}
where 
\begin{eqnarray}
{\cal G}_n(V)\equiv m_n^2 \int_V D_F(m_n |x|) d^4x. \nonumber
\end{eqnarray}
The function ${\cal G}_n(V)$  determines the volume dependence
of the topological susceptibility for large values of $V$. This function 
has a simple behavior\footnote{
For a spherically symmetric volume element with the radius $R$
one can calculate that ${\cal G}_n(V)=1-m_n^2R^2 K_2(m_nR)/2$.}. 
The straightforward calculation yields 
\begin{eqnarray}
{\cal G}_n(0)=0,~~~~~~~~~~~ {\cal G}_n(\infty)=1. \nonumber
\end{eqnarray}
In general, ${\cal G}_n(V)$ is a monotonically increasing function
of the argument.
Its value increases rapidly  from zero  at $V=0$  to almost 
its asymptotic value at some  finite $V$. 
Then, increasing very slowly,  the function  approaches 
the unity as  $V\rightarrow \infty$. 
Relying on these properties one derives 
\begin{eqnarray}
\lim_{V\rightarrow \infty}\chi_{\rm eff} (V)=d -\sum_n f_n^2m_n^2.
\end{eqnarray}
Thus, we see that the topological susceptibility gets additional 
positive subtraction terms in the effective theory (the sum on the
r.h.s.)\footnote[8]{In general, the quantity on the r.h.s. of eq. (9)
is a  nonzero number. However, it was argued in ref. \cite {Samuel}
that the 
topological susceptibility might completely be screened in the 
infinite volume limit 
if instanton interactions are sufficiently strong,  
i.e. $\chi_{\rm eff}(\infty)$ would equal  to zero in that case. 
This condition would yield  a relation 
between the quantity $d$ and parameters of glueballs.
Imposing the condition 
$\chi_{\rm eff}(\infty)=0$ \cite {Samuel} one 
derives $d=\sum_n f_n^2m_n^2$. This relation
is the analog of the Witten-Veneziano formula \cite {Witten} 
\cite {Veneziano} for the $\eta'$ meson mass  (if one
considers full QCD and combines the relation derived above with the 
Witten-Veneziano formula one necessarily needs to take into account
the fact that the value of  $d$ depends on whether it is calculated in 
pure YM theory or in full QCD).  
Thus, the relation $d=\sum_n f_n^2m_n^2$ is a 
phenomenological criterion of the validity of the 
proposal of ref.  \cite {Samuel}. 
That  relation can be tested 
in lattice QCD studies  by measuring $d$ in noninteracting
instanton gas picture of pure YM theory and also by studying  masses and 
decay constants  of the whole tower of pseudoscalar glueball states. 
For the mass and decay constant of the lightest glueball
QCD sum rule results can also be used \cite {Narison}, \cite {FG}.}. 

Finally, using eq. (8)  and   the monotonicity of the function
${\cal G}_n(V)$ one concludes that the topological
susceptibility decreases from its value defined at  relatively small
volumes to  its value reached in the large  volume limit. 
Physically this can be thought
of as following. Suppose we set a sequence  of subvolumes
enclosing some topological charge distribution 
$V_1 <V_2 <V_3  <...< \infty $. 
The result of our discussion is that
$\chi_{\rm eff}(V_1)>\chi_{\rm eff}(V_2)>\chi_{\rm eff}(V_3)>...
\chi_{\rm eff}(\infty)$.
Thus, for smaller volumes one gets larger  values
of  the topological susceptibility.  
One should remember, however, that this picture 
emerges in the confining phase of the theory, i.e., when
$V_1>>\zeta^4$ and 
effective degrees of freedom are colorless excitations.
The interpretation in terms of colored gluons  does not
make sense in that region because of the lack  of asymptotic $in$ and $out$
states for those excitations. 

Recalling that the behavior of the function ${\cal G}_n(V)$
is governed by exponents of the type $e^ {(-m_n R)}$, one 
concludes that 
the effective size  at which $\chi_{\rm eff}(V)$ gets  substantially 
suppressed
is defined by the Compton wavelength of the lightest 
$0^{-+}$ glueball state (eq. (8)). 
In accordance with lattice  
calculations  the lightest pseudoscalar glueball
of pure YM theory is expected to have  mass approximately equal to 
2.3 GeV \cite {UKQCD}. Thus, the effective suppression length scale is 
$L=1/m_{G_0}\simeq 0.09~{\rm fm}$. 

We complete  this subsection by listing  the   main qualitative
conclusions of the discussion presented above.

(i) In the domain of asymptotic freedom, where YM theory
is defined most accurately, the topological susceptibility is 
an increasing function of the argument. 

(ii) In contrast, in the phase where composite, 
colorless excitations are formed   the topological susceptibility
decreases monotonically.

(iii) The suppression length of the topological susceptibility 
is defined by the inverse
mass of the lightest $0^{-+}$ glueball state and equals approximately
$0.09~{\rm fm}$. This length is less  than the 
effective radius of the $0^{-+}$ glueball itself 
(approximately $0.7-1.0~{\rm fm}$ 
\cite {ShuryakGlueballs}, \cite {Chanowitz}). 

An underlying nonperturbative  mechanism which is 
responsible for the formation of the $0^{-+}$  
glueball state, most likely,  should also be responsible 
for the suppression of the topological susceptibility and  vice versa. 
However, the dynamical reason underlying these properties  
is not captured  by the qualitative discussion of this section.
The question why can 
this happen will be addressed below.

Let us notice that the properties listed above 
are in analogy with what happens in the (2+1)-dimensional
Polyakov \cite{Polyakov} model. In the case
of the Polyakov model those features  can be  derived in a rather 
model-independent way \cite {Polyakov}, \cite {Vergeles}.
\vspace{0.3in} \\
{\bf 2.  The $\theta$ Angle and Formation of Glueballs}
\vspace{0.1in}

In this section we study how glueballs can  be formed
in the vacuum of YM theory. In order to address this question let us
first recall how quark containing hadrons are formed in QCD \cite
{CDG79}, \cite {Shuryak78}.
It was found in ref. \cite {CDG77} that nonperturbative fluctuations
lower the value of the vacuum energy density in QCD: If the ground state
energy density for a  perturbative vacuum was zero, then
instantons lower it 
yielding a  negative value \cite {CDG77}. 
When colored quarks  are submerged in that vacuum 
the QCD ground state  responds to the insertion of the quarks by suppressing
the instanton density in a small domain around the quarks \cite {CDG79}. 
In other words, quarks, being submerged  in the YM vacuum,  
yield  a  positive energy density which in the domain
around the quarks partially compensates the existed negative 
ground state energy density. 
The size of that domain
is determined by the dynamics of nonperturbative QCD
\cite {CDG79}.  Hence, if one takes the value of the vacuum 
energy density inside the quark containing domain 
and subtracts the value of the vacuum energy density outside the domain
one would  be left with a positive energy density 
excess in the interior of the domain.

Having a  positive energy excess inside of some region  means that there
should be  an inward pressure acting on each small volume element  of that 
domain. In other words, the outside region with
the negative energy density  produces a pressure on the 
quark containing domain tending to squeeze its volume down to zero.
The quark confinement emerges in this  picture as an effect
of the complicated  structure of the QCD ground state. 
This serves as a derivation of the bag model for
hadrons \cite {bag1}, \cite {bag2} 
from fundamental principles of nonperturbative QCD \cite {CDG79}. 

In this section we show that the same phenomenon might occur 
in pure YM theory. The crucial difference from the previous case  
is the existence of 
a  purely gluonic domain with a positive energy density 
excess. That positive energy density  can be  
provided by the $\theta$ term.  We show below that
the positive  energy density in the interior of the domain is 
proportional to the value of $\theta^2$ multiplied by the value 
of the topological susceptibility. Since the topological
susceptibility
is screened outside of some region, this naturally yields  a  
compact region  of space with a positive vacuum energy density excess
inside.  We show that 
this domain can hadronize  forming  a  YM glueball state.  

Let us start with the action for Yang-Mills theory 
with the CP  violating $\theta$ term. 
In this section we  work in Minkowski space-time.

We  decompose the action $S$ into the usual CP  even  $S^{(+)}$, and 
CP  odd   $S^{(-)}$,  parts.  $S=S^{(+)}+S^{(-)}$, where
\begin{eqnarray}
S^{(+)}=\int  d^4x \left (-{1\over 4
g^2}G^a_{\mu\nu}G^a_{\mu\nu}\right ),~~~~~
S^{(-)}=\theta \int d^4x \left ({1\over 32 \pi^2}
G^a_{\mu\nu}{\tilde G}^a_{\mu\nu}\right ).
\end{eqnarray}
The total energy density of this system ${\cal E}$ is the 
sum  of the energy densities of the CP even part  ${\cal E}^{(+)}$,
and CP odd part ${\cal E}^{(-)}$, i.e. ${\cal E}={\cal E}^{(+)}+
{\cal E}^{(-)}$.  

Let us consider the CP even  part of the action. 
As we mentioned above, nonperturbative contributions 
yield a negative vacuum energy density \cite {CDG77}. 
The total energy density of
the CP even part   is a sum of the negative ground 
state energy density ${\cal E}_{\rm vac}^{(+)}$ and the energy density 
of perturbations  about that ground state 
${\cal E}_{\rm pert}^{(+)}$ 
\begin{eqnarray}
{\cal E}^{(+)}={\cal E}_{\rm vac}^{(+)}+{\cal E}_{\rm pert}^{(+)}.
\end{eqnarray}
Suppose we   start  with no 
perturbations being  excited, i.e.  put ${\cal E}_{\rm pert}^{(+)}=0$.
Then, 
$ {\cal E}^{(+)}={\cal E}_{\rm vac}^{(+)}=
{1\over 4}\langle 0|\Theta^{(+)}_{\mu\mu} |0  \rangle =
{\beta (\alpha_s) \over 4 \alpha_s^2} 
\langle 0|  {1\over 32 \pi^2} G_{\rho\tau}^2 |0  \rangle\simeq
-(0.250~{\rm GeV})^4$ \cite {SVZ}.  Here, $\Theta^{(+)}_{\mu\mu} $
stands for the anomalous trace of the energy-momentum tensor corresponding to 
$S^{(+)}$ and perturbative contributions to the gluon condensate 
are subtracted. 

Let us now address the question what is the contribution
of the CP odd part of the action to the total energy
density of the whole system given in eq. (10). 

It is  convenient   to   introduce 
a new variable by
rewriting  the expression for the topological charge
density\footnote[1]
{Though in Minkowski space-time  $Q$ does not have the  meaning  of
the topological charge density and, moreover, differs from Euclidean
definition
of the topological charge by $i$, we formally keep that
name and letter for simplicity.}  $Q$ 
in terms of a four-form field  $F^{\mu\nu\alpha\beta}$
\begin{eqnarray}
Q = {1\over
4!}\varepsilon_{\mu\nu\alpha\beta}F^{\mu\nu\alpha\beta}, \nonumber
\end{eqnarray}
where the four-form field $F^{\mu\nu\alpha\beta}$ is the field
strength for the three-form potential $C_{\mu\nu\alpha}$ 
\begin{eqnarray}
F_{\mu\nu\alpha\beta}=\partial_\mu C_{\nu\alpha\beta}-
\partial_\nu C_{\mu\alpha\beta}-\partial_\alpha C_{\nu\mu\beta}-
\partial_\beta C_{\nu\alpha\mu}.  \nonumber 
\end{eqnarray}
The $C_{\mu\nu\alpha}$ field  itself is defined as a   
composite operator of colored gluon fields $A^a_\mu$
\begin{eqnarray}
C_{\mu\nu\alpha}={1\over 16 \pi^2}(A^a_\mu 
{\overline {\partial}}_\nu
A^a_\alpha-A^a_\nu {\overline {\partial}}_\mu A^a_\alpha-A^a_\alpha 
{\overline {\partial}}_\nu
A^a_\mu+ 2 f_{abc}A^a_\mu A^b_\nu  A^c_\alpha), \nonumber
\end{eqnarray}
with $f_{abc}$ being  structure constants of the corresponding $SU(N_c)$
gauge group.  The right-left  derivative in this expression acts  as 
$A{\overline {\partial}}B\equiv A (\partial B)-(\partial A) B $.

The topological charge 
density can also be expressed through the Chern-Simons
current $K_\mu $ as   $Q=\partial_\mu K_\mu $. Using 
this expression one can deduce  the relation between the Chern-Simons
current and the three-form potential $C_{\nu\alpha\beta}$, these two
quantities are  Hodge dual to each other
$K^{\mu}={1\over 3!}\varepsilon^{\mu\nu\alpha\beta}C_{\nu\alpha\beta}$.

Let us rewrite 
the CP  odd part of the action  in terms of the 
three-form potential $C_{\nu\alpha\beta}$. For the further 
convenience the integration over  space-time will be restricted
to a finite, not yet specified domain denoted by ${\cal M }$ 
\begin{eqnarray}
S^{(-)}=\theta \int_{\cal M}Q d^4x =-{\theta \over 4!}\int_{\cal M}
F_{\mu\nu\alpha\beta}~dx^\mu \wedge dx^\nu \wedge dx^\alpha 
\wedge dx^\beta \equiv -\theta \int_{\cal M}F, \nonumber
\end{eqnarray}
where  the following differential four-form was introduced
\begin{eqnarray}
F\equiv {1\over 4!}F_{\mu\nu\alpha\beta}~dx^\mu 
\wedge dx^\nu \wedge dx^\alpha 
\wedge dx^\beta . \nonumber
\end{eqnarray}
In terms of  differential forms and an exterior derivative $d$
the equations above formally simplify\footnote[9]{We apologize for 
using the same letter $d$ for an exterior derivative utilized  in this
Section  and the quantity $d$, which has to do with the instanton charge
density, defined  in the previous Section.}.
Indeed,  $F=dC$, where 
$C\equiv {1\over 3!}C_{\nu\alpha\beta}~
dx^\nu \wedge dx^\alpha \wedge dx^\beta $ and the expression for 
$S^{(-)}$ reads as 
\begin{eqnarray}
S^{(-)}=-\theta \int_{\cal M}F=-\theta \int_{\partial {\cal M}}C=
-{\theta \over 3!} \int_{\partial {\cal M}}C_{\nu\alpha\beta}~
dx^\nu \wedge dx^\alpha \wedge dx^\beta .
\end{eqnarray}
In the last equation we used the Stokes  theorem assuming that
the boundary ${\partial {\cal M}}$ enclosing the  domain 
${\cal M}$ is an orientable smooth surface. 
Speaking in terms of the $C_{\nu\alpha\beta}$
field, the nonzero value of the $\theta$
angle corresponds in Minkowski space 
to the nonzero coupling of the $C_{\nu\alpha\beta}$
field to the  boundary manifold ${\partial {\cal M}}$. That  coupling
is gauge invariant,  although 
the $C_{\nu\alpha\beta}$ field itself is not a gauge invariant
quantity. Indeed, if the gauge transformation parameter is 
denoted by $\Lambda^a$, then the three-form field transforms as 
$C_{\nu\alpha\beta}\rightarrow
C_{\nu\alpha\beta}+\partial_\nu \Lambda_{\alpha\beta}-
\partial_\alpha \Lambda_{\nu\beta}-\partial_\beta
\Lambda_{\alpha\nu}$,
where $\Lambda_{\alpha\beta}\propto A_\alpha ^a\partial_\beta
\Lambda^a - A_\beta  ^a\partial_\alpha \Lambda^a$. However, it 
is easy to  check that the expression for the 
field strength $F_{\mu\nu\alpha\beta}$ 
is gauge invariant. Since  the coupling of the three-form field 
to the boundary can be expressed in terms of the 
$F_{\mu\nu\alpha\beta}$ field (as in eq. (12)), 
then that  coupling is also gauge
invariant and can lead to some physically observable results.  
The same conclusion could  be drawn  without referring to the 
field strength. The gauge variation of the last 
expression in eq. (12) is zero for any smooth closed surface 
which does not enclose  any field singularities.
  
It was noticed some time ago
\cite {Luscher} that the $C_{\nu\alpha\beta}$  field 
propagates in the bulk
of the domain  ${\cal M}$ if the topological susceptibility
is nonzero in that domain. 
This becomes more evident if  one recalls 
the notion  of the Kogut-Susskind (KS) pole \cite {KogutSusskind} in the
correlator of two Chern-Simons currents. 
We briefly present  those arguments.

Consider the correlator of the vacuum topological susceptibility
at a nonzero momentum. 
The topological charge density $Q$ 
is the derivative of the Chern-Simons current  $Q=\partial_\mu K_\mu$.
One can  substitute this definition back into the expression
for the correlator of the topological susceptibility.
In that way one  discovers  that $\chi $ is defined 
as the zero momentum limit of the correlator of two Chern-Simons
currents multiplied by two momenta\footnote{The multiplier $-i$
appears in the definition of the topological susceptibility in
Minkowski space-time. There are some delicate
issues regarding the  definition of the correlator of the vacuum topological 
susceptibility. If one defines $\chi$ as a second derivative of the 
partition function w.r.t. the $\theta$ angle, 
then some contact term  appears  in that expression  \cite {Witten}. 
Likewise, a special care is needed while 
treating the covariant $T$ product in
eqs. (1) and (13)  when this last is taken to be the definition
of $\chi$ \cite {Crewther}.  One should add a contact term 
(given in the Appendix of ref. [26]) to the r.h.s. of eq. (1)
in order to define  $\chi$ as a second derivative of the 
vacuum energy  w.r.t. the $\theta$ angle [26,41] (this contact
term can effectively be included in eq. (6) by redefining the value of
the positive constant $d$).} 
\begin{eqnarray}
\chi =i\lim_{q\rightarrow 0}q^\mu q^\nu \int e^{iqx}\langle 0|T
K_\mu (x) K_\nu (0)|0\rangle d^4x.
\end{eqnarray}  
The only way 
for this  expression to be nonzero is to claim that
the correlator of two Chern-Simons  currents develops a pole
as the momentum goes to zero.
This  is called the Kogut-Susskind pole \cite {KogutSusskind}.

Knowing that the correlator of two Chern-Simons  currents 
has a pole, one can use the 
relation between the Chern-Simons current and the three-form 
$C_{\nu\alpha\beta}$  field 
and   conclude that the $C_{\nu\alpha\beta}$ field also has a nonzero
Coulomb  propagator \cite {Luscher}. 
Thus, the $C_{\nu\alpha\beta}$ field behaves as  a 
massless collective excitation transferring  
a long range interaction  \cite {Luscher}. 

Let us summarize  briefly the results of the  discussion given above.
Following ref. \cite{Luscher} we established that the three-form
field  $C_{\nu\alpha\beta}$  propagates in the 
bulk transferring a long-range Coulomb interaction. The exact propagator 
of this field is  of the Coulomb type and is proportional to 
the value of the vacuum topological susceptibility.

We also saw that the CP  odd term in the action of Yang-Mills theory 
can be expressed as a coupling of the three-form composite field 
$C_{\nu\alpha\beta}$ to the boundary manifold. Hence, the three-form
field $C_{\nu\alpha\beta}$ being a free field in the bulk 
actually does couple to the boundary surface. 

All the properties mentioned above can  be 
summarized in the following effective action  for the $C_{\nu\alpha\beta}$
field: 
\begin{eqnarray}
S^{(-)}_{eff}=-{1\over 2\cdot 4!\chi (V_{\cal M} ) }\int_{{\cal M}} 
F^2_{\mu\nu\alpha\beta}~d^4x
-{\theta \over 3!} \int_{\partial {\cal M}}C_{\nu\alpha\beta}
~dx^\nu \wedge dx^\alpha \wedge dx^\beta .
\end{eqnarray}
The first term in this expression 
yields  the correct Coulomb propagator
for the three-form $C_{\nu\alpha\beta}$ field. The second term is just the 
usual CP  odd 
$\theta$ term of the initial YM action. 
$V_{\cal M}$ denotes the three-volume of the domain ${\cal M}$.
Notice that higher derivative terms are neglected in this action as
they are suppressed by momenta of massless three-form field. 

Our next step is to study the effective action  given in eq. (14)
\footnote[8]{One should  notice that the  action (14)
is not an  effective action in the Wilsonian sense. It is rather
related to the generating functional of one-particle-irreducible
diagrams of the composite field. The effective action in eq. (14) 
is not to be quantized  and loop diagrams of that action 
are not to be taken into
account in calculating  higher order Green's functions. 
The analogous effective action for
the CP even part of the theory was constructed in 
refs. \cite {Schechter},  \cite{MigdalShifman}.}.
In particular,  we will calculate the ground state 
energy of the system using the effective action (14). 
Before we turn to that calculation let us mention
that Maxwell's equations for a  free four-form potential 
$F_{\mu\nu\alpha\beta}$ yield only a constant solution in 
$(3+1)$-dimensional  
space-time \cite {Aurilia}. The reason  is the following.
A four-form potential has only one independent degree of freedom
in four-dimensional space-time, let us call it $\Sigma$. Then, 
the four Maxwell's equations written in terms of the $\Sigma$ field
ensure that this field is independent of all four space-time 
coordinates. Hence, the solution can only be a space-time constant.
Thus, the $F_{\mu\nu\alpha\beta}$ field propagates 
no dynamical degrees of freedom\footnote{The exception is when that  field
couples to other fields.}. 
However, this field can be  
responsible for the existence of a positive vacuum energy density 
in different models of Quantum Field Theory (see ref. \cite {Townsend}).
Thus, studying classical equations of motion for the 
$F_{\mu\nu\alpha\beta}$ field one can  determine  
the value of the ground state energy given by configurations of 
$F_{\mu\nu\alpha\beta}$.   
We are going to solve explicitly classical equations  of motion
for the effective action (14). Then,  
the energy density associated with those solutions will be calculated.

Let us start with the equations of motion. Taking the variation of the 
action (14) with respect to the $C_{\nu\alpha\beta}$  field one gets 
\begin{eqnarray}
\partial^\mu F_{\mu\nu\alpha\beta}(z)=\theta \chi(V_{\cal M})
\int_{\partial {\cal M}} \delta^{(4)}(z-x)
~dx_\nu \wedge dx_\alpha \wedge dx_\beta . \nonumber
\end{eqnarray}
It has been shown  in ref. \cite {Aurilia} that this kind of
equations can  be solved exactly in four-dimensional space-time.
The solution is the sum of a particular solution of the 
inhomogeneous equation and a  general solution of the homogeneous
equation 
\begin{eqnarray}
F_{\mu\nu\alpha\beta}(z)=\theta \chi(V_{\cal M} )
\int_{{\cal M}}\delta^{(4)}(z-x)
~dx_\mu \wedge  dx_\nu \wedge dx_\alpha \wedge dx_\beta +h~ 
\varepsilon_{\mu\nu\alpha\beta}.
\nonumber
\end{eqnarray}
The integration constant $h$, if nonzero,  induces an  additional 
CP violation beyond the existed  $\theta$ angle \cite
{Townsend}. Since we would like to avoid to have an extra CP 
violating term  we set $h=0$.
Simplifying the previous equation one finds that the classical
solution is a nonzero constant tensor density inside 
of the domain  ${\cal M}$
\begin{eqnarray}
F_{\mu\nu\alpha\beta}=-\theta \chi(V_{\cal M} )
\varepsilon_{\mu\nu\alpha\beta},
\end{eqnarray}
and is zero, $F_{\mu\nu\alpha\beta}=0$, outside of ${\cal M}$.

As a next step let us 
compute the vacuum energy associated with 
the solution given in eq. (15). The density of the 
energy-momentum tensor for the CP 
odd  sector of  the theory can be written as 
\begin{eqnarray}
\Theta^{(-)}_{\mu\nu}=-{1\over 3! \chi (V_{\cal M})   }\left (
F_{\mu\alpha\beta\tau}F_{\nu}^{~\alpha\beta\tau}-{1\over 8}
g_{\mu\nu} F^2_{\rho\alpha\beta\tau}\right ). \nonumber
\end{eqnarray}
Using the expression (15)
one calculates the corresponding 
energy density\footnote{One can use either 
${\cal E} ={1\over 4}\Theta^\mu_{~\mu}$
or ${\cal E}=\Theta_{00}.$ } ${\cal E}^{(-)}$ 
\begin{eqnarray}
{\cal E}^{(-)}={1\over 2}\theta^2 \chi (V_{\cal M}). \nonumber
\end{eqnarray}
Since the $F_{\mu\nu\alpha\beta}$ field does not propagate 
dynamical degrees of freedom the expression above is the 
total energy density of the system given by the action  (14). 
The crucial  thing  about
this   energy density is that 
it is a positive quantity 
proportional to  $\theta^2$  multiplied by  the value of the 
topological susceptibility\footnote[3]{One  might wonder
whether the same result is obtained if one treats
$\theta$ not as a constant multiplying  $Q$ in the Lagrangian,
but as the phase that the states acquire under a topologically
non-trivial gauge transformations. These two ways of presenting the 
$\theta$ dependence are equivalent.
Thus, results of our discussion should be equivalent in both cases. 
The key observation 
is that if  $\theta$ is not entering the
Lagrangian,  the arbitrary integration constant 
in eq. (15) has to be nonzero. It should be chosen 
in a way that would guarantee a  
proper $\theta$ dependence of the VEV of the topological charge density.
That would leave the results of our discussion without modifications.}. 

We learned in the preceding section that
the magnitude of the topological susceptibility depends on the 
value of the subvolume in which it is calculated, and also, most
importantly, it is screened by nonperturbative effects of YM theory
outside of some finite subvolume element.

So far we treated the domain  ${\cal M}$ as some arbitrary
volume. Let us now suppose that  ${\cal M}$ is the  
subvolume outside of which the topological susceptibility is 
screened\footnote[2]{In this discussion we 
assume   that the domain has a more or less definite 
boundary, or in other words,  that there is a  
narrow interval where the topological
susceptibility drops in magnitude from the value $d$ to 
$d-\sum_n f_n^2m_n^2$. That  assumption does  not seem to be far from
the reality if one recalls  the 
behavior of the function ${\cal G}_n(V)$ (see Section 1).}.   
Thus, the topological susceptibility 
is given  by the quantity $d$ inside of the volume ${\cal M}$
and by the difference   $d-\sum_n f_n^2m_n^2$ outside of the volume. 
As we mentioned above, the  
difference  $d-\sum_n f_n^2m_n^2$
turns  to  zero if interactions between instantons are  
sufficiently strong \cite {Samuel}. However, this is not guaranteed
in general. Trying to deal with the most general case, we assume here that 
$d-\sum_n f_n^2m_n^2$ is some number not necessarily equal to zero.
Clearly, our result presented below will also be applicable to  the case when 
the topological susceptibility is zero outside of the volume  ${\cal M}$
and $d=\sum_n f_n^2m_n^2$. In accordance with our 
previous calculations, 
the vacuum energy density  inside 
of  the domain ${\cal M}$  is   
${\cal E}_{\rm inside}={\cal E}^{(+)}_{\rm vac}+
{1\over 2}\theta^2 d $. Let us now turn to the 
vacuum energy density outside of the volume  ${\cal M}$. 
As we  assumed,   the topological susceptibility 
is not necessarily zero outside of ${\cal M}$. Hence, the three-form field
can propagate in that region too and 
$F_{\mu\nu\alpha\beta}\neq 0$ outside 
of ${\cal M}$ in the general case. 
As a result, there exists some nonzero vacuum energy density outside of 
${\cal M}$. In analogy with the previous case one 
derives  the 
following expression for the vacuum energy density 
${\cal E}_{\rm outside}={\cal E}^{(+)}_{\rm vac}+{1\over 2}\theta^2
(d-\sum_n f_n^2m_n^2)$. This value is less than the energy 
density  inside of the volume ${\cal M}$.
Thus,  there is a positive
energy density  excess inside of ${\cal M}$. 
The expression for
the corresponding  energy excess in the subvolume takes the form
\begin{eqnarray}
\Delta E={1\over 2}\theta^2 \Delta \chi V_{\cal M}.
\end{eqnarray}
Here  $\Delta \chi\equiv \chi_{\rm inside}-\chi_{\rm outside}= 
\sum_n f_n^2m_n^2$ is the 
difference between the topological susceptibility defined inside 
and outside of the subvolume. As we mentioned, eq. (16) is also valid
in the particular case when the topological susceptibility equals to zero 
outside of the volume ${\cal M}$. In that case  
$\Delta \chi=\sum_n f_n^2m_n^2 =d$ and the energy difference in
eq. (16) coincides with the energy of the CP odd 
part inside of the volume ${\cal M}$. 

There are two basic questions to be elucidated  here. 
First of all why would any distinguished subvolume 
exist in the YM vacuum?
The reason, as we already mentioned above,  is the screening of the 
topological susceptibility.
This naturally provides finite domains
in the YM  vacuum with the positive vacuum energy excess 
in accordance with eq. (16).

The second question is what happens 
with this  finite volume  if it is allowed to 
flow freely  to a stationary state. 
The system will  tend  to minimize the  energy given in
eq. (16).  The expansion is not an 
energetically allowed process. 
An alternative possibility  for the system is to squeeze  
its volume  down as much as it is possible. In that case  
the r.h.s. of eq. (16)  would  be decreasing.
In other words, 
there should be an inward  pressure acting on the system and tending
to squeeze its volume down. That  
pressure is  due to  the  positive  difference between the 
energy densities  in the interior and exterior of the domain. 

Hence, one concludes that 
the system will tend  to minimize its energy by squeezing its volume 
down, or decreasing $V_{\cal M}$ in eq. (16). 

At first glance such a system is unstable 
and should collapse to a point. However, that  would be a  wrong
conclusion.  The point  is that we did not yet
take into account perturbations of YM fields  
which should  get  excited inside of $V_{\cal M}$
while  the system is shrinking its volume down. Those 
excitations could stabilize the system.
The energy of those excitations, identified in
eq. (11) as ${\cal E}^{(+)}_{\rm pert}$, 
provides
an additional contribution  to the total energy. 
Anticipating the 
results of the next section we  present the expression for the 
total energy inside of the domain ${\cal M}$ 
\begin{eqnarray}
\Delta E_{\rm total}={1\over 2}\theta^2 \Delta \chi  V_{\cal M}+
{({\rm positive~number})\over  V_{\cal M} ^{1/3}}, 
\end{eqnarray}
where the first term is related to the CP odd part  of the 
initial action and the second one is the contribution of perturbations
of the CP even part. 

The structure of this equation allows one to minimize 
the quantity  $\Delta E_{\rm total}(V_{\cal M} )$  with respect to  
$V_{\cal M} $
and find an  optimal value of the three-volume occupied by the system. 

We treat this physical system as a model for a pure YM 
glueball state. 

The spectrum and some properties of that system are
studied below.
\vspace{0.3in} \\
{\bf 3. Yang-Mills vs. Closed Membrane  Spectrum}
\vspace{0.2in} 

It was shown in the previous section that the screened topological 
susceptibility  leads to a  positive 
energy density excess inside of some finite volume. The system
tends to minimize that volume. 
The compression of the  volume
continues  until some  YM states are exited
inside of that domain. Those states have nonzero energy, i.e.,  
$\langle  {\cal H}_{YM} \rangle \ne 0$, 
where  ${\cal H}_{YM}$ is the Hamiltonian density
of YM theory (the CP even part). 
We mentioned already that the phenomenon described above  
is related to the fact that $\chi$ is screened.
On the other hand,  we saw  that
the effects responsible for the screening of the topological charge 
should also be responsible for the formation of the $0^{-+}$ glueball state.
Hence, it is reasonable  to identify those finite volume YM excitations
with physical $0^{-+}$ glueball states. 

In this section we 
study the spectrum of the physical YM Hamiltonian in a  finite
volume. 
Under some approximations, elucidated  below, 
the spectrum of YM theory resembles that  
of a  closed  bosonic membrane with the topology
of a sphere or torus \cite{Hoppe1}, \cite{Hoppe2}. 
One can use that  analogy  to derive the relation between the
spectrum  of  YM theory and  that  of a closed bosonic membrane.
Using that  relation and  calculating the spectrum for 
a  closed spherical bosonic membrane we predict  masses
for  two lightest pseudoscalar YM glueball states. 
\vspace{0.2in} \\
{\it 3.1.   Studying the spectrum of Yang-Mills theory}
\vspace{0.1in}

Let us start with  the physical Hamiltonian density
of YM theory. 
In order to stress  the approximations we make   we  present the  
brief discussion of the Hamiltonian formalism of the theory
(for detailed
discussions see the textbooks \cite {Dirac}, \cite {FaddeevSlavnov} and 
\cite {GitmanTiutin}).

One starts  with the Lagrangian density of  pure YM theory 
in Minkowski space-time
\begin{eqnarray}
{\cal L} =-{1\over 4
g^2}G^a_{\mu\nu}G^a_{\mu\nu}. \nonumber
\end{eqnarray}
The canonically conjugate momentum is defined as the derivative of the
Lagrangian density w. r. t.  the time derivative of the canonical 
coordinate and is given by
$$
P^a_i=-{1\over g^2}G_{0i}^a.
$$
The Hamiltonian density takes the form 
\begin{eqnarray}
{\cal H}_0={g^2\over 2}P^a_iP^a_i+ {1\over 4g^2}G_{ij}^a G_{ij}^a+
A_0^a(D_iP_i)^a, \nonumber \\ \nonumber i,j=1,2,3.~~~~~~~~~~~~~~~~~ 
\end{eqnarray}
This is not the physical Hamiltonian density yet.
There are extra degrees of freedom in this expression. 
The existence of those extra variables  
is  related to the gauge invariance
of the theory. 

The Lagrangian density does not contain 
time derivatives of the $A_0$
field. As a result, the following primary constraint  appears
$ P^a_0=0. $ 

Introduce \cite {Dirac} the so called  ``total'' 
Hamiltonian $H_T(t)\equiv \int d^3x ({\cal H}_0+\lambda^a
P^a_0)$,
where $\lambda^a(x)$  denotes  a  Lagrange multiplier. 
Time evolution of a physical quantity  is given by the Poisson
brackets of $H_T$ and the quantity itself. Thus, one needs to set 
conventions for the Poisson brackets. 
For any two (bosonic) functionals  $A$ and $B$ we use 
the following expression:
$$
\{A, B\}\equiv \int d^3z \left ( {\delta A\over \delta q(z)}
{\delta B\over \delta p(z)}-{\delta B\over \delta q(z)}
{\delta A\over \delta p(z)}\right ),
$$
where $q$ and $p$ denote canonical coordinates and momenta respectively. 
Using this definition 
one finds that the conservation of the primary constraint  
$\{P^a_0(x, t),H_T(t)\}=0$,
leads one to the secondary constraint in the form of the Gauss's  law
$D^{ab}_iP^b_i=0$. One can also check that the 
conservation of the Gauss's law is identically 
satisfied and no further constraints are  produced at this stage. 

We are going to work in 
the axial gauge $A_3^a=0$. Requiring the conservation
of the gauge  condition  $\{A_3^a(x,t),H_T(t)\}=0$, one derives
the 
additional secondary constraint  $g^2P^a_3-\partial_3 A_0^a=0$.
Finally, the conservation of that   constraint leads to the 
equation   for determination of the Lagrange multiplier 
$\partial_3 \lambda^a(x)+\partial_3 \partial_iA_i^a(x)-f^{abc}A_0^b
\partial_3 A_0^c=0$. Thus, the whole system
of gauge conditions and constraints can be summarized as 
\begin{eqnarray}
\Phi_1^a=P^a_0,~~~~~~~~~~~~~~~~~~~\Phi_2^a=D^{ab}_iP^b_i, \nonumber \\
\Phi_3^a=A_3^a,~~~~~~~~~~~\Phi_4^a=g^2P^a_3-\partial_3
A_0^a. \nonumber
\end{eqnarray}
The physical Hamiltonian in the axial gauge can be written in terms
of the following physical variables $P_m^a$ and $A_m^a$, where 
$m=1,2$ \cite {GitmanTiutin}.
In general, the straightforward procedure 
implies  the elimination  of  all  nonphysical variables
by solving (wherever it is  possible)  constraint equations and 
substituting  those expressions
back into the formula  for the Hamiltonian. 
In most cases the result is a complicated   nonlocal expression for the
Hamiltonian. There  is a formally simpler  way to follow, however.
One can solve only some  part of  the constraint equations 
keeping  the rest  of  the constraints
unsolved and  allowing  some of  nonphysical variables to be present
in the Hamiltonian. Then, the physical system is 
defined by that  Hamiltonian accompanied by unsolved constraint
equations imposed on the physical states of  the Fock space.
For our purposes we found it convenient to follow that  way. 
Using the conditions $\Phi_1^a=0,~\Phi_3^a=0$, and $\Phi_4^a=0$ 
the Hamiltonian density can be rewritten as 
\begin{eqnarray}
{\cal H}_{YM}={g^2\over 2}P^a_mP^a_m+ {1\over 4g^2}G_{mn}^a G_{mn}^a+
{1\over 2g^2}(\partial_3 A_0^a)^2+{1\over 2g^2}(\partial_3 A_m^a)^2,
\end{eqnarray}
where $m,n=1,2$ and the  expression contains the physical variables
$A_m^a$ and $P_m^a$ along with the nonphysical $A_0^a$. 
The  constraint which is still left  relates  
$A_0^a$ to the 
physical  variables 
\begin{eqnarray}
\partial_3^2A_0^a+g^2(D_m P_m)^a=0.
\end{eqnarray}
Thus, the system is defined  by the Hamiltonian density (18)
and the  constraint (19). 

Let us now turn to the discussion of the spectrum of the system (18-19)
which is placed in a finite three-volume denoted by $V_{\cal M}$.  
Calculating the spectrum  we are going to keep only  
``slow'' modes, i.e. the modes with zero momenta but a nonzero energy.
All the ``fast'' modes with nonzero momenta can be thought of as being  
integrated out. The net result  of the corrections due to the 
``fast'' modes is just a  perturbative splitting of the energy levels 
determined by the ``slow''  modes 
(for a detailed discussion see ref. \cite {LuscherNP})
\footnote[7]{The crucial  point  in this discussion is that
the spectrum is calculated in a small volume limit. As we mentioned  
in Section 1, this corresponds to the weak coupling approximation. 
As a result,  corrections due to the ``fast'' modes are of order
${g^{2/3}\over 4\pi}=({\alpha_s\over 16\pi^2})^{1/3}$  \cite
{LuscherNP}  and can be neglected in the leading approximation.}.  
Adopting that approximation  one can  drop all the spatial derivatives
in the expression for the Hamiltonian and constraint equation 
assuming that all the canonical variables
depend on the time variable only. 

Let us turn to the Hamiltonian  instead of the Hamiltonian density. 
Dropping  all the spatial derivatives one writes down
$$
H_{YM}= {g^{2} V_{\cal M} \over 2} 
P^a_mP^a_m+ { V_{\cal M} \over 4 g^2}(f^{abc}A_m^bA_n^c)^2.
$$
It is convenient to perform  the following rescaling of the 
canonical variables  
$$
A_m \rightarrow {g^{2/3}\over V_{\cal M}^{1/3}}A_m~~~~ {\rm  and }~~~~~
P_m \rightarrow {1\over g^{2/3} V_{\cal M} ^{2/3} }P_m.
$$ 
The new, rescaled variables are dimensionless. 
In terms of these  variables 
the expression for the Hamiltonian takes the form
\begin{eqnarray}
H_{YM}= {g^{2/3}\over V_{\cal M} ^{1/3}} \left [ {1\over 2}
P^a_mP^a_m+ {1\over 4}(f^{abc}A_m^bA_n^c)^2 \right ],
\end{eqnarray}
and the constraint equation is given as follows:  
\begin{eqnarray}
f^{abc}A_m^bP_m^c=0,~~~~~~m=1,2.
\end{eqnarray}
This is the system which defines the spectrum\footnote[4]{
The operator in (20) acts on  functionals of the canonical variable
while the momentum operator is defined as $P_m=-i{\delta\over\delta
A^m}$.}.  The first thing
to notice  is that the potential in eq. (20) has flat directions. 
Thus, one would  expect  a continuous spectrum 
without a mass gap. However, it was  proved in ref. \cite {Simon}
(see also \cite {LuscherNP})
that in the quantum theory, contrary to the
naive classical expectation, the operator
defined in eq. (20) has only discrete  positive eigenvalues. As a result,
the following expression for the spectrum  emerges:
\begin{eqnarray}
E^{(+)}_{\rm pert}\equiv \langle H_{YM}  \rangle ={\cal E}^{(+)}_{\rm pert}
V_{\cal M}={ g^{2/3}\times ({\rm positive~~number})
\over V_{\cal M}^{1/3}}. \nonumber 
\end{eqnarray}
This expression was used earlier in eq. (17). The exact calculation of
the positive numbers  occurring in the expression above is a complicated
problem of YM theory. However, as it will be shown below, 
one can use some analogies and calculate
the spectrum explicitly. We turn now 
to that  discussion. 
\vspace{0.2in}\\ 
{\it 3.2.  The Membrane Matrix Model} 
\vspace{0.1in} 

It was shown some time ago \cite {Hoppe1}, \cite {Hoppe2} that 
the Hamiltonian
of a closed bosonic membrane in the light-cone gauge 
can  be reduced to  the 
form given in eq. (20). 

The variables substituting the gauge fields  
in that case occur as  coefficients of the harmonic expansion 
of the spatial coordinates on the membrane world surface. 
The two Hamiltonians, one for the membrane and the other one 
given in eq. (20)  formally look  similar. 

The  YM theory constraint (21) also has an  
analog  in the case of the closed
membrane theory. The constraint in that  case is related to  the residual
reparametrization
invariance of the membrane action which is still left 
in the light-cone gauge.

Below we discuss briefly the membrane action
and the way  it  reduces  to the form given in eq. (20).
Then we deduce  the matching condition relating the spectrum of
the  closed bosonic membrane to the spectrum  of YM theory. 
The matching condition allows one to obtain the 
spectrum of YM theory by calculating the spectrum of
the closed bosonic membrane.   

We present below only the basic features  of the membrane Hamiltonian
construction in the light-cone gauge. 
For details  we refer to the original papers \cite {Hoppe1}, 
\cite {Hoppe2}. 

The membrane action in flat Minkowski
space-time  can be written as
\begin{eqnarray}
S_m=-T \int d^3 \sigma \sqrt{|{\rm det} g_{ij}|},
\end{eqnarray}
where $T$ is the membrane tension, 
the constant with the dimensionality  of mass cubed; $\sigma_0, \sigma_1,
\sigma_2 $ are the coordinates on the membrane world volume;
$g_{ij}$ denote the components of  the induced metric in the 
membrane world volume
\begin{eqnarray}
g_{ij}(\sigma)\equiv {\partial X^\mu (\sigma )  \over \partial \sigma^i}
{\partial X_\mu (\sigma)  \over \partial \sigma^j},
\end{eqnarray}
where $X_\mu, ~~\mu=0,1,2,3,$ are the space-time coordinates.

The membrane action is reparametrization invariant.  
Thus, in accordance with the Noether second theorem 
not all of the variables in the action are independent  
(as in gauge theories).
One should carry out the gauge fixing procedure. 
It is convenient to introduce the light-cone coordinates
\begin{eqnarray}
X^{\pm}={1\over \sqrt 2} (X^3\pm X^0), \nonumber
\end{eqnarray}
and choose the light-cone gauge
\begin{eqnarray}
X^{+}(\sigma)=X^{+}(0)+\sigma_0. \nonumber
\end{eqnarray}
The light-cone gauge does not completely fix the 
gauge freedom of the membrane action \footnote {As opposed 
to the case of a string action where 
in the light-cone gauge no  freedom is left  \cite {GSW}.}.  
As a result, there still is 
a residual local  invariance left.  Hence, one should expect to have
the Hamiltonian of the theory accompanied by a  constraint equation.
The detailed discussion and the construction 
of the Hamiltonian is given in refs. 
\cite {Hoppe1}, \cite {Hoppe2}. We present the final result here.
The expressions for the mass squared operator and the constraint 
can be written as follows: 
\begin{eqnarray}
{M^2\over 2}= \left [ {1\over 2}
{\cal P}^a_m{\cal P}^a_m+ {T^2 \over 4}(g^{abc}X_m^bX_n^c)^2 \right ],
\end{eqnarray}
\begin{eqnarray}
{g} ^{abc}X_m^b{\cal P}_m^c=0~~~~~~m=1,2. \nonumber
\end{eqnarray}
The canonical coordinates and momenta are the functions 
of the time variable only. The coordinates $X_m^a$ in this expression 
are the coefficients of the harmonic expansion 
of the space-time coordinates $X_m$ on the surface of the membrane.
For example,  if the membrane has  the 
topology of a sphere, then
the harmonic expansion mentioned above 
is just the expansion of the space-time coordinates 
in the basis of spherical functions 
$$
X_m(\sigma)=\sum _{a=1}^{\infty} X_m^a Y^a(\sigma_1,\sigma_2 )
,~~~~a=1,2...\infty, 
$$
where $Y^a(\sigma)$'s are the harmonic functions on the  sphere.

If the membrane has  the topology of a sphere or torus 
the  harmonic functions $Y^a(\sigma)$ form 
a representation of the Lie algebra
of the SU($\infty$) gauge group \cite {Hoppe1} 
\cite {Hoppe2}\footnote[8]{The SU($\infty$) 
group (and its Lie algebra) should be understood as a 
limit of the SU(N) group at $N\rightarrow \infty$.}.
Thus, the  $SU(\infty)$ gauge group appears due to the 
reparametrization invariance of the membrane 
action\footnote[4]{The supersymmetric version of the 
membrane matrix model is used for the formulation
of the M theory in the infinite momentum frame \cite {BFSS}.}.

The expression (24) 
resembles the Hamiltonian
of the YM system in the approximation given in eq. (20) and in the 
$N_c\rightarrow \infty$ limit. The constraint equations 
in the two cases are also similar. 

In order to make use of this analogy  let us 
perform the following rescaling of the canonical variables
\footnote{One can check that the rescaling procedures
we perform lead to dimensionless canonical
coordinates and momenta which satisfy   the commutation
relation with the unity on its r.h.s.}
\begin{eqnarray}
{\cal P} \rightarrow T^{1/3} {\cal P} ~~~~~{\rm and }~~~~~~ 
X\rightarrow T^{-1/3}X.
\nonumber 
\end{eqnarray}
The new canonical variables are dimensionless. The expression for
the mass squared  operator in terms of those variables takes the form
\begin{eqnarray}
{M^2\over 2}= T^{2/3}\left [ {1\over 2}
{\cal P}^a_m {\cal P}^a_m+ {1 \over 4}(g^{abc}X_m^bX_n^c)^2 \right ].
\end{eqnarray}
Thus, one concludes that the spectrum of a closed  bosonic
spherical membrane is determined by the same differential operator
as the one for YM theory in the large $N_c$ limit\footnote[3]{In both
cases these differential operators act on the functionals
of the canonical coordinates. Those functionals are annihilated 
by the constraint equations. The boundary conditions
will be fixed later during the actual calculations.}. 
Matching this expression  with  eq.  (20)
one  finds  the relation  between  the spectrum of  
YM theory in a finite volume and the spectrum of the 
closed bosonic membrane
\begin{eqnarray}
E_n^{YM}={g^{2/3}\over V_3^{1/3}} {M_n^2 \over 2 T^{2/3}},
\end{eqnarray}
where $M_n$'s are the mass eigenvalues defined by the operator 
given in  eq. (25). 
The complimentary constraint equations 
acting on the physical states ensure  that the 
physical eigenfunctionals
of the Hamiltonians in both eqs. (20) and (25) are  the functionals of
``colorless'' (gauge invariant) variables only.  
Indeed, in both cases the constraint
equations  (the Gauss's law and its membrane counterpart)
serve as the generators of the ``gauge'' 
transformations of the initial system. Since those 
generators are supposed to annihilate any physical state 
(imposed as the Gauss's law annihilating a state), 
then all the physical eigenfunctionals should be  gauge invariant. 
 
In the next subsection we  calculate $M_n^2$ for a 
closed membrane with the  topology of a sphere and using the matching
condition (26) deduce  the energy levels for the YM
theory excitations identifying them  with the 
pseudoscalar glueballs.
\vspace{0.2in} \\ 
{\it 3.3. Calculating the Membrane Spectrum}  
\vspace{0.1in} 

We start with a  closed spherical  membrane. The space-time 
coordinates on the membrane world surface are given as
\begin{eqnarray}
X_\mu =(t,~~ r(t){\rm sin}\theta {\rm cos}\varphi,
~~r(t){\rm sin}\theta {\rm sin} \varphi, ~~
r(t){\rm cos} \theta ), \nonumber  \\ \nonumber 
0\le \theta \le \pi, ~~~~~~~~~~0\le \varphi \le 2\pi,
\end{eqnarray}
where $r(t)$ is the time dependent radius of the 
membrane. 

The induced metric on the membrane worldvolume has the following  
nonzero components:
$$
g_{tt}=1-{\dot r}^2(t),~~~ g_{\theta\theta}= -r^2(t),
~~~~  g_{\varphi\varphi}= -r^2(t) {\rm sin}^2\theta. 
$$
The action functional
for the  membrane takes the form 
\begin{eqnarray}
S=-T \int dt d\theta d\varphi r^2(t) {\rm sin}\theta \sqrt {1-{\dot r}^2(t)}.
\nonumber
\end{eqnarray}
Thus, the Lagrangian can be written as  follows:
\begin{eqnarray}
L(t)=- 4\pi T r^2(t) \sqrt {1-{\dot r}^2(t)}. \nonumber
\end{eqnarray}
Calculating the canonically conjugate momentum
\begin{eqnarray}
P={\partial L(t)\over \partial {\dot r}}=4 \pi T {r^2(t) {\dot r}(t)
\over  \sqrt {1-{\dot r}^2(t)}}, \nonumber
\end{eqnarray}
one derives the Hamiltonian for the spherical membrane\footnote[9]{
The Hamiltonian looks similar to  the one for a  relativistic particle 
with the time-dependent mass $m(t)= 4\pi T r^2(t)$ and describes
the pulsation of the spherical membrane.}
\begin{eqnarray}
H=\sqrt {P^2+ 16 \pi^2 T^2 r^4}. \nonumber
\end{eqnarray}
As we mentioned above  we are looking for 
the mass squared operator for the membrane  (see eq. (25)).
Thus,  we need to solve the following Schr{\" o}dinger equation
\begin{eqnarray}
M_n^2 \Psi (r)= \left (-{d^2\over dr^2}+16 \pi^2 T^2 r^4 \right )\Psi (r),
\nonumber
\end{eqnarray}
with the boundary conditions  $\Psi (\infty)=0$ and $\Psi (0)=0$.

It is useful to turn to the dimensionless variable $z$ 
defined as 
$$
z\equiv r (16 \pi^2 T^2)^{1/6}.
$$
In terms of  $z$ the Schr{\" o}dinger equation takes the
form
\begin{eqnarray}
{M_n^2 \over (16 \pi^2 T^2)^{1/3}} \Psi (z)= 
\left (-{d^2\over dz^2}+z^4 \right )\Psi (z).
\end{eqnarray}
The Schr{\"o}dinger equation with the quartic potential has been
extensively studied in the literature (for a review see ref. 
\cite {Spectrum}). 
The results of  numerical calculations  of  the first ten eigenvalues
can be found in ref. \cite {Spectrum}. Those calculations 
 are  usually 
done for  
the potential defined on the whole z axis.
 In our case z is defined on the 
positive semiaxis only. Thus, only the odd parity solutions 
are relevant for the present  case. Those solutions have
nodes at $z=0$ and satisfy the boundary conditions  
$\Psi (0)=0$ and $\Psi (\infty)=0$.

Here, we present only the first two parity odd eigenvalues of eq. (27)
\begin{eqnarray}
{M_0^2 \over (16 \pi^2 T^2)^{1/3}}=2.393 644, \nonumber
\\ \nonumber
{M_1^2 \over (16 \pi^2 T^2)^{1/3}}=7.335 730.
\end{eqnarray}
Using these  expressions and the matching condition (26) one calculates
the first two energy levels  for the YM system in the finite volume
\begin{eqnarray}
E_{YM}^0={g^{2/3}\over V_{\cal M}^{1/3} } { (16 \pi^2 )^{1/3}\over 2} 
2.393 644, \\  
E_{YM}^1={g^{2/3}\over V_{\cal M}^{1/3} } { (16 \pi^2 )^{1/3}\over 2} 
7.335 730.
\end{eqnarray}
The numerical values for the energy levels are determined by the 
strong coupling constant  $g$ and also by the volume of the
domain ${\cal M}$. The strong coupling constant is supposed to be taken
at the scale appropriate for corresponding  glueballs.  

Let us take eqs.  (28) and (29) 
and substitute them into eq. (17). This leads to 
the expression for the total energy inside of the finite volume we are
discussing
\begin{eqnarray}
\Delta E(V_{\cal M})={1\over 2} \theta^2 \Delta \chi  V_{\cal
M}+{u^2_n\over V_{\cal M}^{1/3}},
\end{eqnarray}
 where in accordance with eqs. (28-29)
\begin{eqnarray}
u^2_0=g^{2/3}(2 \pi^2 )^{1/3}2.393 644,~~~{\rm and}~~~
u^2_1=g^{2/3}(2 \pi^2 )^{1/3}7.335 730.
\end{eqnarray}
The expression (30) can be minimized w.r.t. the 
value of the three-volume $V_{\cal M}$. We denote the optimal value
for the volume by $ {\bar V}$, hence 
${d\Delta E(V_{\cal M})\over d V_{\cal M}}|_{\bar V}=0$.
Using this condition
and taking the derivative of eq. (30) one finds
\begin{eqnarray}
{1\over 2} \theta^2 \Delta \chi \simeq {1\over 3}
{u^2_n\over {\bar V} ^{4/3}},
\end{eqnarray}
and the value of the total energy for the optimal volume
\begin{eqnarray}
\Delta E({\bar V}_n)\equiv m_n \approx 
{4\over 3}{u^2_n\over {\bar V}_n^{1/3}}.
\end{eqnarray}
Here, we denote by $m_n$ the mass of the corresponding $n$'th glueball
and by ${\bar V}_n$ the corresponding optimal value of  the volume element.
Thus, knowing the value of the strong coupling constant
at the scale appropriate for the lightest glueballs 
(which is about $1.5-2.5~{\rm GeV}$) and also knowing the value of the 
effective size of the YM $0^{-+}$ glueball state one can predict the 
value of its mass by means of eq. (33).

We present below the results of calculations
for three different values of the strong  coupling 
constant $\alpha_s$. 
The reasonable estimate for the lightest pseudoscalar 
glueball radius is $R_0=0.7-1.0 ~{\rm fm}$ \cite {Chanowitz}, 
\cite {ShuryakGlueballs}. 
The size of the second excited glueball state $R_1$ is not known. 
However, using eqs. (32) and (31) one can estimate 
that $R_1\approx 1.3 R_0=(0.9-1.3)~{\rm fm}$.  
The results of numerical calculations of glueball masses 
for those values of the coupling constant and radii 
are presented below.  
\begin{eqnarray}
{\underline {\alpha_s=0.3}}~~~~~~~~~~~~~~~~~~~~~~~~~~~~~~~~~~~~
~~~~~~~
\nonumber  \\ \nonumber
R_0=0.7~{\rm fm},~~~~~m_0=2340~{\rm MeV},~~~~~~R_1=0.91~{\rm fm},     
~~~~~~m_1=5520~{\rm MeV}. \\ \nonumber
R_0=0.8~{\rm fm},~~~~~m_0=2050~{\rm MeV},~~~~~~R_1=1.04~{\rm fm},
~~~~~~m_1=4830~{\rm MeV}. \\ \nonumber
R_0=0.9~{\rm fm},~~~~~m_0=1820~{\rm MeV},~~~~~~R_1=1.17~{\rm fm},
~~~~~~m_1=4300~{\rm MeV}. \\ \nonumber
R_0=1.0~{\rm fm},~~~~~m_0=1640~{\rm MeV},~~~~~~R_1=1.30~{\rm fm},
~~~~~~m_1=3870~{\rm MeV}. \\ \nonumber
{\underline {\alpha_s=0.35}}~~~~~~~~~~~~~~~~~~~~~~~~~~~~~~~~~~~~~~
~~~~~
 \\ \nonumber
R_0=0.7~{\rm fm},~~~~~m_0=2470~{\rm MeV},~~~~~~R_1=0.91~{\rm fm},
~~~~~~m_1=5800~{\rm MeV}. \\ \nonumber
R_0=0.8~{\rm fm},~~~~~m_0=2160~{\rm MeV},~~~~~~R_1=1.04~{\rm fm},
~~~~~~m_1=5090~{\rm MeV}. \\ \nonumber
R_0=0.9~{\rm fm},~~~~~m_0=1920~{\rm MeV},~~~~~~R_1=1.17~{\rm fm},
~~~~~~m_1=4520~{\rm MeV}. \\ \nonumber
R_0=1.0~{\rm fm},~~~~~m_0=1730~{\rm MeV},~~~~~~R_1=1.30~{\rm fm},
~~~~~~m_1=4070~{\rm MeV}. \\ \nonumber
{\underline {\alpha_s=0.4}}~~~~~~~~~~~~~~~~~~~~~~~~~~~~~~~~~~~~~~~
~~~~~
 \\ \nonumber
R_0=0.7~{\rm fm},~~~~~m_0=2580~{\rm MeV},~~~~~~R_1=0.91~{\rm fm},
~~~~~~m_1=6080~{\rm MeV}. \\ \nonumber
R_0=0.8~{\rm fm},~~~~~m_0=2260~{\rm MeV},~~~~~~R_1=1.04~{\rm fm},
~~~~~~m_1=5320~{\rm MeV}. \\ \nonumber
R_0=0.9~{\rm fm},~~~~~m_0=2010~{\rm MeV},~~~~~~R_1=1.17~{\rm fm},
~~~~~~m_1=4730~{\rm MeV}. \\ \nonumber
R_0=1.0~{\rm fm},~~~~~m_0=1805~{\rm MeV},~~~~~~R_1=1.30~{\rm fm},
~~~~~~m_1=4260~{\rm MeV}.
\\ \nonumber
\end{eqnarray}
These predictions can be 
compared with the result  of the lattice calculation for the lightest 
pseudoscalar glueball mass  $m_0=2.3\pm 0.2~{\rm GeV}$ \cite {UKQCD}. 
We should stress that the masses presented above 
give  just the large $N_c$ approximation  to the actual values. 
We regard these numbers as
reasonable estimates for the pseudoscalar glueball masses. 

Let us now discuss an interesting consequence of eq. (32).
If one knew  the effective size of the glueball and also the 
value of $\Delta \chi $, then one would be 
able (using  eq. (32)) to calculate  the value of the 
$\theta$ parameter 
\begin{eqnarray}
\theta^2 \approx {2\over 3}  {u_n^2\over {\bar V}_n^{4/3} \Delta
\chi}. \nonumber
\end{eqnarray}
In general,  the value of 
$\Delta \chi $ is not known. However, in order to get an order of magnitude
estimate for $\theta$ one can crudely approximate $\Delta \chi $
by the lightest glueball contribution $ f_0^2m_0^2$ multiplied by
the number of $0^{-+}$ glueballs in the spectrum 
of the model (let us call
that number  $N$):
$\Delta \chi \approx Nf_0^2m_0^2\approx
N (200~{\rm MeV})^4$ \cite {Narison}, \cite {FG}. 
Then, if  $\alpha_s=0.3$ and the  
lightest glueball  radius $R_0=0.8~{\rm fm}$
the $\theta$ parameter should be equal to $\theta \approx 6/ \sqrt {N}$.
One can also estimate the magnitude  of $\theta$ 
for different values of the radius. Generically, if the value of $N$ is not 
too large, the magnitude  of $\theta$ is of order of the unity or so.

Some comments are in order here. First of all the estimate  for the 
$\theta$ parameter presented above appears  as a result  
of the physical
picture of the glueball formation discussed  in this work.
However, the method of modeling  the glueball spectrum 
by means of the membrane Hamiltonian does not depend on a particular
mechanism of the formation of glueballs. Indeed, whatever the mechanism 
of the formation is, the glueball can always be regarded in some extent
as a close domain  of space where  the YM excitations are confined  
and the spectrum of which is  determined by the YM 
Hamiltonian given in eq. (20). 

The second comment concerns the strong CP problem.
In this work we deal  with pure YM theory. No light quark degrees of 
freedom were  
included. The large value of the $\theta$ parameter 
that  we derived  should somehow be neutralized 
when quark degrees of freedom are taken into account. 

There are some  possibilities
for that. We list below three of them.

In full QCD the parameter which defines the magnitude of the 
strong CP violation is the sum of the $ \theta $ angle 
used in this work and the  phase 
of the determinant of the quark mass matrix, $arg~det M$. 
It is possible that those two contributions compensate each other
and the strong CP violation, being present in pure YM theory,
does  not appear in full QCD. This might lead to an interesting pattern
of mixing between the pseudoscalar glueball and the $\eta'$ meson,
when the pseudoscalar glueball being present in pure YM model,
might not appear in full QCD as a separate state.  

The second possibility is realized if  one has 
a massless quark in the model. In that case the $\theta$ 
dependence can be eliminated from the QCD Lagrangian by an appropriate chiral
rotation of that quark field. 
What happens with the glueball state  in full QCD  remains to be studied.

Finally,  one can  argue (using the results of refs. \cite
{Screening}, \cite {Samuel}) that in full QCD
the $\eta'$ meson, mediating interactions  between topologically charged
objects, provides a sufficient (from the experimental point of view) 
screening of the 
topological susceptibility even in the massive theory. 
In terms of eq. (9) that can  be understood by including the 
$\eta'$ contribution on the r.h.s. and deducing a Witten-Veneziano
type relation.  More detailed studies of full QCD 
are  needed in order to determine which of the above
scenarios (if any) can actually be realized. 
\vspace{0.3in} \\
{\bf Discussions}
\vspace{0.1in} 

In this paper we studied some properties
of the YM vacuum which should  be responsible 
for the formation of the $0^{-+}$ glueball states. 

The properties of the correlator of the vacuum 
topological susceptibility as a function of the volume element $V$
are discussed.
In the weak coupling (small volume) approximation it is 
an increasing function of the argument.  Increasing the volume
continuously  the theory passes through 
a crossover region after which it 
should  be regarded  as a strongly correlated one. 
Above the  crossover region  the topological susceptibility
becomes a rapidly decreasing function of the argument
and reaches its asymptotic value (not necessarily zero)
in the large volume limit. 
Thus, the value of the vacuum topological susceptibility 
is screened if the strong coupling regime of the theory
is considered.

It is shown that the presence of the $\theta$ angle in the 
theory along with the screening phenomenon can lead to the 
formation of a glueball state. An  important ingredient of that
scenario is the existence of the three-form composite field
propagating the Coulomb-like interaction. 

The spectrum of the YM Hamiltonian resembles 
in the zero momentum approximation  the spectrum  of a closed 
bosonic membrane. Using that  analogy and calculating the spectrum of a
closed bosonic  membrane  we estimate  the masses of glueballs in the
large $N_c$ limit. 
The result for the lightest $0^{-+}$ 
glueball is  in  agreement with the lattice prediction. 
We also predict the mass of the next-to-lightest 
glueball. This result can be checked in future lattice
calculations. In general, our approach allows us to compute  the
mass of any heavier glueball state (if such a state exists).
The method of calculation of the spectrum is in general independent 
of the mechanism by which glueballs are formed in YM theory and 
the YM vs. the membrane Hamiltonian analogy utilized for that calculation
can always be applied. 

Notice that   
the large $N_c$ arguments were not used  while deriving eq. (17). 
The large  $N_c$ approximation 
was  adopted later on  in order  to
calculate the ``positive number'' occuring on the r.h.s. of
eq. (17). Thus, the approach
and equations presented  in this work are not peculiar to the 
$N_c\rightarrow \infty$ limit. They should rather have some wider range of 
validity beyond the large $N_c$ approximation. 
For instance, the second term in eq. (17)
can be thought as a result of the uncertainty 
principle alone.  

There are a number of interesting questions  left out of the discussion in 
the present paper. First of all we did not discuss the fate 
of a scalar glueball. The effective Lagrangian
approach to the $0^{++}$ channel of pure YM theory was developed in 
refs. \cite{Schechter},  \cite {MigdalShifman}. 
One can apply  the YM Hamiltonian 
vs. the membrane Hamiltonian analogy to the calculation of 
the scalar glueball mass too. 
This last would correspond to the lowest parity-even solution 
of the Schr\"odinger equation (27). Hence, the scalar glueball
would emerge  to be  lighter than the pseudoscalar one. This is in
agreement with what is known from various lattice and theoretical 
studies \cite {UKQCD}, 
\cite {West}. However, the mechanism of the formation of the 
scalar glueball can not be  captured by our analysis. 

We did not discuss here how colored degrees of freedom
are confined inside of a finite closed volume. It was rather assumed
that QCD provides this property  by some 
mechanism. Formally,  it was assumed that
the operator in the Gauss's law, being the generator of  
gauge transformations, should annihilate all the physical
states. Thus, all those states 
are supposed to be colorless states by the construction.
In various models of hadrons,
confinement can be warranted by imposing some 
boundary conditions on fields, 
as in the case of the MIT bag model \cite {bag1} or the model of 
ref.  \cite {Gnadic}, or by postulating
some specific  dielectric properties of the vacuum as in the case of
the Friedberg-Lee model \cite {FL}. Some discussions  of these 
issues  from the point of view of QCD can be found in ref. \cite
{CDG79}.   

Finally,  one needs to know what happens
when quark degrees of freedom are also included in the theory. 
In that case the  mixing between the $\eta'$ meson and the glueball
should play an important role (if those two states exist
simultaneously).  Our discussion of the three-form field
in that  respect becomes  crucial. It is known that the $\eta'$
meson couples to the topological charge density, hence it couples
to  the three-form potential too. 
Thus, one can naturally couple  the $\eta'$ meson to 
the glueball by means of  the three-form field.
These and other related 
questions  will  be addressed elsewhere.
\vspace{0.3in} \\
{\bf Acknowledgments}   
\vspace{0.1in}

The author is grateful to G.R. Farrar, M. Porrati and M. Schwetz 
for reading the manuscript and for useful suggestions and discussions.
I wish to thank  T. Banks for valuable  discussions. 
The work was supported by grant No. NSF-PHY-94-23002.  
\vspace{0.3in} \\
{\bf Appendix}   
\vspace{0.1in}

In this appendix we consider the dispersion relation for the 
correlator of the vacuum topological susceptibility in  momentum space.
The space-time  is assumed to be a Euclidean one.
The correlator is defined as in 
eq. (1),  Section 1. Before we go  further let us   specify how  
singularities are handled in eq. (1). The product of
two operators of the topological charge density is singular 
as  $x\rightarrow 0$. The leading perturbative singularity 
at $x\rightarrow 0$ can be calculated 
\begin{eqnarray}
T Q(x) Q(0) \propto {1\over x^8}+O\left ({ln x^2\over x^8} \right ).
\nonumber
\end{eqnarray}
Upon integration in eq. (1) this  expression yields a divergent 
term. 
A simple way to handle the divergence is to
allow a small momentum $k$  to flow through the correlator function
treating  $\chi(V)$ as a  zero momentum limit 
of the corresponding momentum-dependent renormalized Green's function 
$$
\chi(V)=
\lim_{k^2\rightarrow 0} \chi^{ren} (k^2, V)=
\lim_{k^2\rightarrow 0} ~\left [ \int_V e^{ikx}
\langle 0|T Q(x) Q(0)|0\rangle d^4x \right ]^{ren}, 
$$ 
where $k$ is the regularizing 
momentum. This  relation implies
that the limiting procedure  is supposed to be
carried out  after the integration and renormalization 
of the divergent parts 
are  already done in  momentum space.  In what follows we 
adopt this prescription.

Another type of divergence  occurring in eq. (1) is related
to the $x\rightarrow \infty $ limit. In that limit
$$
\langle 0|T Q(x) Q(0)|0\rangle \rightarrow \langle 0|Q|0\rangle 
\langle 0|Q|0\rangle.
$$
Supposing that generically the VEV of the topological charge density
might not be zero in a CP violating model,
one gets the divergence in eq. (1) as  
$V\rightarrow \infty $.
In order to eliminate this divergence one can work with the 
subtracted correlator. This amounts to  saying  that the actual integrand in
eq. (1) is the function with the following subtraction
$$
\langle 0|T Q(x) Q(0)|0\rangle -\langle 0|Q|0\rangle  \langle 0|Q|0\rangle.
$$
The subtracted function goes to zero in the  $x\rightarrow \infty$ limit. 
The coordinate-independent 
subtraction term  does not affect  our analysis and 
was  dropped  for simplicity in Section 1.  It will also be omitted below. 

In what follows we show that continuum contributions
vanish  in the limit   $k^2\rightarrow 0$.
 
The dispersion relation for the correlator of the vacuum topological 
susceptibility in momentum space can be written as
$$
\chi(k^2)=\chi(0)+\chi'(0)~k^2 +{k^4\over \pi}\int_{m^2_{G_0}}^{\infty}
{\rho(s) ds \over s^2(s-k^2)},    \eqno ({\rm A} 1)
$$
where $\rho(s)\equiv {\rm Im}~ \chi(s+i\epsilon)$.  
The correlator at zero momentum is denoted by $\chi(0)$. The quantity 
$\chi'(0)$ stands for the derivative
of the correlator w.r.t. $k^2$ at  $k^2=0$.

In order to make the integral convergent, 
and also to account for the correct asymptotic behavior of the
correlation function at $k^2\rightarrow \infty$, we have  introduced 
the subtraction terms in the dispersion relation (A1). 

The dispersion relation in the form given is eq. (A1) is not
convenient  for our purposes. In the limit $k^2\rightarrow 0$
it turns into a  trivial identity. One  needs to rewrite (A1) in  a form
similar to the one given in  eq. (6).
For this purpose  let us use the following identity:
\begin{eqnarray}
{k^4 \over s^2(s-k^2)}={1\over s-k^2}-{1\over s}-{k^2\over s^2}.
\nonumber
\end{eqnarray} 
Substituting this formula into eq. (A1)  
one rewrites the dispersion relation in the following form:
$$
\chi(k^2)=d_0+b_0~k^2 +{1\over \pi}\int_{m^2_{G_0}}^{\infty}
{\rho(s) ds \over s-k^2}, \eqno ({\rm A}2)
$$
where
\begin{eqnarray}
d_0\equiv \chi(0) -{1\over \pi}\int_{m^2_{G_0}}^{\infty}
{\rho(s) ds \over s}, ~~~~~
b_0\equiv \chi'(0) -{1\over \pi}\int_{m^2_{G_0}}^{\infty}
{\rho(s) ds \over s^2}.
\nonumber
\end{eqnarray} 
The form of the relation given in eq. (A2) is very 
formal one. The constants $d_0$, $b_0$ and the integral on the r.h.s.
are  divergent quantities. When these terms are put 
together all divergences cancel and the whole expression 
is finite. The divergences mentioned above
are related to  perturbative contributions to the 
spectral density.
Thus, it is convenient
to separate  nonperturbative and perturbative
terms.  We found it  useful  to apply  
the decomposition usually adopted in QCD sum rule calculations 
\cite {SVZ}.  One decomposes
the expression for the spectral density  
\begin{eqnarray}
\rho(s)=\rho^{np}(s)+\rho^{pt}(s)\vartheta(s-s_0),
\nonumber
\end{eqnarray} 
where the superscripts $''np''$ and $''pt''$ denote  
nonperturbative and perturbative terms, respectively. Here $\vartheta$
denotes  the step function. 
The constant $s_0$ sets the continuum threshold (or the duality
interval) \cite {SVZ} and by the definition $s_0>m^2_{G_0}$. 
It is  assumed in this approach that  resonance
contributions are defined by the nonperturbative part of the spectral
density.  One also supposes that  due to 
asymptotic freedom   
continuum contributions above the continuum 
threshold can be approximated by leading perturbative terms \cite{SVZ}. 

Let us make the same formal decomposition for 
the quantities $d_0$ and $b_0$ 
\begin{eqnarray}
d_0=d+d^{pt},~~~~~d^{pt}=-{1\over \pi}\int_{s_0}^{\infty}
{\rho^{pt}(s) ds \over s}, 
\nonumber
\end{eqnarray} 
\begin{eqnarray}
b_0=b+b^{pt},~~~~~b^{pt}=-{1\over \pi}\int_{s_0}^{\infty}
{\rho^{pt}(s) ds \over s^2}.  
\nonumber
\end{eqnarray} 
Here $d$ and $b$  are  the quantities determined
by the complicated vacuum structure of YM theory. 
As we mentioned already, in the weak coupling approximation 
with noninteracting
instantons   
$d$ is defined as the value of the topological susceptibility
of a dilute instanton gas in the large  volume limit of 
pure YM theory.
The quantity $d$ appears  in eqs. (5) and (6-8)  in the text.
Finally, using all the expressions given above one derives 
$$
\chi(k^2)=d+b~ k^2 +{1\over \pi}\int_{m^2_{G_0}}^{\infty}
{\rho^{np}(s) ds \over s-k^2}+{k^4\over \pi}\int_{s_0}^{\infty}
{\rho^{pt}(s) ds \over s^2(s-k^2)}. \eqno ({\rm A}3)
$$
We should notice here that eqs. (A3) and (A1) differ from each other
by some formal redefinitions. Moreover, eq. (A3) is written
adopting some particular scheme of separation between 
perturbative and nonperturbative contributions. That  procedure
is not unambiguous. In that respect, eq. (A3) should
be regarded as an expression defined within the framework of  
the particular prescription outlined above. 

Now one can use the fact that the quantity $k^2$ is a  
regularizing momentum. Thus, one can assume
that $k^2$ is very small,  so  that the condition $s_0>>k^2$
is readily satisfied. The 
last integral on the r.h.s of eq. (A3) can be expanded 
in a power series of the ratio 
$k^2/s_0$ (since that integral is convergent). 
Performing the   expansion, and then 
Fourier transforming  eq. (A3) with the weight ${1\over (2 \pi)^4}$,
one derives
the expression for the correlator $\langle 0|T Q(x) Q(0)|0\rangle $
in the following form\footnote[5]{We use the following normalization 
for the delta function: $\delta^{(4)}(x)={1\over (2 \pi)^4}
\int_{-\infty}^{+\infty}e^{ikx}d^4k$ and $\delta^{(4)}(k)=
\int_{-\infty}^{+\infty}e^{-ikx}d^4x$.}
$$
\langle 0|T Q(x) Q(0)|0\rangle=d~\delta^{(4)}(x)-b~
\partial^2\delta^{(4)}(x)+{1\over \pi}\int_{m^2_{G_0}}^{\infty}
\rho^{np}(s) D_F(\sqrt{s} |x|)~ds+
$$
$$
+{1\over \pi}\sum _{n=2}^{\infty}(-1)^n
\int_{s_0}^{\infty}
{\rho^{pt}(s)\over s} \left ({\partial^2 \over s} \right )^n
\delta^{(4)}(x)~ds.   \eqno ({\rm A}4)
$$
Eq. (A4) is a general  form of the expression given in eq. (6) 
in Section 1. 
In order to reproduce the sum on the r.h.s.
of eq. (6) one needs to make the following substitution in eq. (A4)
$$\rho^{np}(s)=-\pi\sum_n f^2_n m_n^4 ~\delta (s-m_n^2).$$
Some terms on  the r.h.s.
of eq. (A4) with  derivatives of the Dirac delta function
yield vanishing contributions  upon integration
in eq. (1). For that  reason  those  derivative containing 
terms were omitted in eq. (6).

\end{document}